\documentstyle[12pt]{article}

\begin{document}
\title{$2^{++}$ glueball}
\author{Bing An Li\\
Department of Physics and Astronomy, University of Kentucky\\
Lexington, KY 40506, USA}

\maketitle
\begin{abstract}
The mixing between the $f_2(1270)$, the $f_2(1525)$, and the $2^{++}$ glueball is determined and tested. 
The mass and the hadronic decay widths of the $G_2$ and the branching ratio $B(J/\psi\rightarrow\gamma G_2)$ are predicted. 
\end{abstract}
\newpage
\section{Introduction}
The glueball states are solutions of nonperturbative QCD. The study of $2^{++}$ glueball has a long history. The MIT bag model 
predicts \(m(2^{++}) = 1.29 \textrm{GeV}\) [1]. In Ref. [2] the mass of the $2^{++}$ glueball state is predicted in the range
of $1.45 - 1.87 \textrm{GeV}$. In Refs. [2,3] it is argued that there are glueball components in the $f_2(1270)$ and the $f_2(1525)$  
mesons. In Ref. [4] the smallness of the ratio of the helicity amplitude $y={T_2\over T_0}$ of the decay 
$J/\psi\rightarrow \gamma f_2(1270)$ is explained by that the meson $f_2(1270)$ contains substantial component of $2^{++}$ 
glueball. 
%$B(\Upsilon(1S)\rightarrow\gamma f_2(1270))$ is about one order of magnitude smaller than 
%$B(J/\psi\rightarrow\gamma f_2(1270))$ [6]. 
There are many studies on $2^{++}$ glueball [5].
The mass of the $2^{++}$ glueball has been calculated by quenched lattice QCD to be about 2.39 $\textrm{GeV}$ [6].
On the other hand, many $2^{++}$ isoscalar states have been discovered [7]: $f_2(1430),\;f_2(1565),\;f_2(1640),\;f_2(1810),\;f_2(1910),\;
f_2(1950),\;f_2(2010),\;f_2(2150)...$. Some of them are radial excitations of the $f_2(1270),\;f'(1525)$.
It is very possible that a $2^{++}$ glueball is among them. Of course, a physical glueball state contains both $|q\bar{q}>$
and $|gg>$ states.

In this paper the mixing of the $f_2(1270),\;f'(1525)$ and the $2^{++}$ glueball is studied and tested.
The mass and the hadronic decay width of the $2^{++}$ glueball
$G_2$ and the branching ratio of $J/\psi\rightarrow\gamma G_2$ are predicted. These results can be used to identify
the $2^{++}$ glueball.

\section{Mixing of the $f_2(1270),\;f'(1525)$ and the $G_2$ glueball and the mass of the $G_2$}
According to QCD, two expansions are applied in this study. One is the chiral expansion, the expansion of the 
current quark mass, $m_q$, and the second is the $N_C$ expansion. In QCD it is known $g^2\sim {1\over N_C}$, where 
g is the coupling constant of gluons and quarks.
The $f_2(1270),\;f'(1525)$ and the $G_2$ glueball are the eigen states of the mass matrix of the $f_8,\;f_0$ and the 
pure $2^{++}$ glueball $g_2$, where $f_8,\;f_0$ are the $2^{++}$ octet and singlet states. The mass matrix is expressed as
\begin{equation}
\left( \begin{array}{c c c}
                  m_1\;\Delta_1\;\;\;\Delta_2\\
                  \Delta_1\;\;\;m_2\;\;\Delta_3\\
                  \Delta_2\;\;\Delta_3\;\;m_3
                  \end{array} \right),  
\end{equation}
where \(m_1=m^2_{f_8},\;m_2=m^2_{f_0},\;m_3=m^2_{g_2}\), and \(\Delta_1=<f_8|m^2|f_0|>,\;
\Delta_2=<f_8|m^2|g_2>,\;\Delta_3=<f_0|m^2|g_2>\). From the quark model, we obtain
\begin{eqnarray}
m_1 = {1\over3}(4m^2_{K^*} - m^2_{a_2}),\nonumber \\
m_2 = {1\over3}(2m^2_{K^*} + m^2_{a_2}).
\end{eqnarray}
Using the two expansions, we have
\begin{eqnarray}
\Delta_1 \sim O(m_q),\;\;
\Delta_2 \sim O(m_q {1\over N_C}),\;\;
\Delta_3 \sim O({1\over N_C})
\end{eqnarray}
The masses of the $f_8$ and $f_0$ (2) are up to $O(m_q)$. The study presented in this paper is 
up to either $O(m_q)$ or $O({1\over N_C})$.
Obviously, the $\Delta_2$ is at higher order in the two expansions and it can be ignored
\begin{equation}
\Delta_2 = 0.
\end{equation}
$m_3,\;\Delta_1,\;\Delta_3$ are the three undetermined parameters. The $m^2_{f_2}$
and $m^2_{f'_2}$ are taken as inputs. Therefore, one more input is required.

The branching ratios of $f'_2(1525)\rightarrow K\bar{K},\;\pi\pi$ are listed [9] as
\begin{eqnarray}
B(f'_2(1525)\rightarrow K\bar{K}) = (88.7 \pm 2.2)\%,\nonumber \\
B(f'_2(1525)\rightarrow \pi\pi) = (8.2 \pm 1.5)\times 10^{-3}.
\end{eqnarray}
The $B(f'_2(1525)\rightarrow K\bar{K})$ is larger than the $B(f'_2(1525)\rightarrow \pi\pi)$ by two order of magnitudes.
On the other hand, both are d-wave decays. The phase space of the $\pi\pi$ channel is much larger than
the one of the $K\bar{K}$ channel
\[\frac{(1-{4m^2_\pi\over m^2_{f'_2}})^{{5\over2}}}{(1-{4m^2_K\over m^2_{f'_2}})^{{5\over2}}} = 3.61.\]
Therefore, the magnitude of the amplitude of the $f'_2(1525)\rightarrow K\bar{K}$ is about 20 times of the one 
of the $f'_2(1525)\rightarrow \pi\pi$. The physical state of the $f'_2(1525)$ contains both the $q\bar{q}$ and the gluon-gluon
components. It is reasonable to assume that the $\pi\pi$ is from the gluon-gluon component of the $f'_2(1525)$.
Therefore, the $q\bar{q}$ component is dominated by the $s\bar{s}$. The physical state of the $f'_2(1525)$ is expressed as
\begin{eqnarray}
f'_2(1525) = a_2 f_8 + b_2 f_0 + c_2 g,\\
f_8 = {1\over \sqrt{3}}(u\bar{u} + d\bar{d} - 2s\bar{s}), \\
f_0 = {1\over \sqrt{3}}(u\bar{u} + d\bar{d} + s\bar{s}).
\end{eqnarray}
The $s\bar{s}$ dominance in the $f'_2(1525)$ leads to
\begin{equation}
a_2 = - b_2.
\end{equation}
Eq. (9) is another input for determining the mixing.

Now the mass of the physical glueball state $G_2$ and the mixing of the $f_2,\;f'_2,\;G_2$ 
can be determined.
After input the values of the $m^2_{f_8},\;m^2_{f_0},\;m^2_{f_2},\;m^2_{f'_2}$ the three eigen equations are obtained
\begin{eqnarray}
m^2_{G_2} = m_3 + 0.1326,\\
m^2_{G_2} = 1.2345 + 7.5643(\Delta^2_1 + \Delta^2_3),\\
7.5643(\Delta^2_1 + \Delta^2_3) \Delta^2_1 - 1.7604 \Delta^2_1 -0.7186\Delta^2_3 + 0.08447 = 0.
\end{eqnarray}
Using the Eqs. (9,10-12), it is determined
\begin{equation}
\Delta_1 = m_1 - m^2_{f'_2} = -0.1819\;\textrm{GeV}^2.
\end{equation}
Substituting Eq. (13) into Eqs. (11-12), we obtain
\begin{eqnarray}
\Delta_3 = 0.07368\;\textrm{GeV}^2,
m_{G_2} = 1.429\; \textrm{GeV},
m_{g_2} = 1.382\;\textrm{GeV}.
\end{eqnarray}
The mass of the physical $2^{++}$ glueball state is predicted. In Ref. [7] a state $I^G(J^{PC}) = 0^+(2^{++})$ $f_2(1430)$ 
which is listed. The value of $m_{G_2}$ predicted in this study
is close to the results presented in Refs. [1,2], but lower than the value obtained by the quenched Lattice QCD [6]. 

The expressions of the three physical $2^{++}$ states are determined from the three eigen equations of the mass matrix (1)
\begin{eqnarray}
f_2(1270) = 0.246 f_8 + 0.7002 f_0 -0.6702 g_2,\\
f'_2(1525) = -0.6421 f_8 + 0.6421 f_0 +0.4189 g_2,\\
G_2 = 0.618 f_8 + 0.3451 f_0 +0.7064 g_2.
\end{eqnarray}
The $G_2$ state contains substantial $q\bar{q}$ components, the glueball component in the $f_2$ is large and in $f'_2$
is not negligible. In this study the $g_2$ is a pure glueball state and the $G_2$ is a new $2^{++}$ state. Without the
pure glueball state $g_2$ the $G_2$ doesn't exist.

\section{The decays $f_2\rightarrow\pi\pi,\;K\bar{K},\;\eta\eta$ and $f'_2\rightarrow K\bar{K},\;\eta\eta$}
As a test of these results (15-17), the decays of $f_2(1270)\rightarrow\pi\pi,\;K\bar{K}$ are studied in the chiral limit.
They are d wave decays and the decay widths are expressed as
\begin{eqnarray}
\Gamma(f_2\rightarrow\pi\pi) = |T|^2 m_{f_2} (a_1 + b_1)^2(1-{4m^2_\pi\over m^2_{f_2}})^{{5\over2}},\\
\Gamma(f_2\rightarrow K\bar{K}) = |T|^2 m_{f_2} {1\over3}(2b_1 - a_1))^2(1-{4m^2_K\over m^2_{f_2}})^{{5\over2}},
\end{eqnarray}
where $a_1$ and $b_1$ are the coefficients of Eq. (15) and \(a_1 = 0.246,\;b_1=0.7002\).
It is known that the pion and the kaon are Goldstone bosons and $m^2_\pi \propto m_u+m_d$ and
$m^2_K\propto m_s+{1\over2}(m_u+m_d)$. In the chiral limit the 
$|T|^2$ is independent of the current quark masses.  
This study predicts
\begin{equation}
\frac{\Gamma(f_2\rightarrow K\bar{K})}{\Gamma(f_2\rightarrow\pi\pi)} = 0.0548. 
\end{equation}
The experimental value of this ratio is [7] \(0.054(1 \pm 0.12)\). Theory agrees with the data very well.

Because of the mixing between $\eta,\;\eta'$ and the $0^{-+}$ glueball [8] the decay mode $\eta\eta$ is more complicated.
However, it is known that the octet component dominates the $\eta$ state. In this study the $\eta$ is taken as an octet
to calculate $\Gamma(f_2\rightarrow\eta\eta)$.  
The decay width of the $f_2\rightarrow\eta\eta$ is derived as
\begin{eqnarray}
\Gamma(f_2\rightarrow\eta\eta) = |T|^2 m_{f_2} {1\over3}(b_1 - a_1))^2(1-{4m^2_{\eta}\over m^2_{f_2}})^{{5\over2}},\\
\frac{\Gamma(f_2\rightarrow\eta\eta)}{\Gamma(f_2\rightarrow K\bar{K})} = 0.055.
\end{eqnarray}
The experimental data [7] is $0.087(1 \pm 0.29)$. As mentioned above in Eq. (22) there are mixing between $\eta,\;\eta'$ and 
the $0^{-+}$ glueball. 
The theoretical result agrees with the experimental data within
the experimental error reasonably well.
Similarly, the ratio
\begin{equation}
\frac{\Gamma(f'_2\rightarrow\eta\eta)}{\Gamma(f'_2\rightarrow K\bar{K})} = 0.29
\end{equation}
is obtained. The data of this ratio are following
\[0.069(1\pm0.17) [9]\;\;0.33(1\pm0.10) [10],\;\; 0.12 (1 \pm 0.23) [7].\]
\section{The decays $G_2\rightarrow\pi\pi,\;K\bar{K},\;\eta\eta$}
Now we need to study the decays of $G_2(1429)\rightarrow\pi\pi,\;K\bar{K},\;\eta\eta$. 
Because of $g^2\sim {1\over N_C}$ the hadronic decays of the glueball
components $|gg>$ of these states are in higher order in the $N_C$ expansion and suppressed. Therefore,
the hadronic decays are the decays of their $q\bar{q}$ components only. 
Eq. (17) shows that the $G_2(1429)$ state
contains large $|q\bar{q}>$ components. The hadronic decay width of the $G_2(1429)$ state won't be small.
It is reasonable to assume that in the chiral limit the $|T|^2$ in Eqs. (18,19,21) 
are about the same.  
This assumption can be tested by calculating the $\Gamma(f'_2\rightarrow K\bar{K})$ by inputting the $\Gamma(f_2\rightarrow K\bar{K})$.
Replacing the quantities of the $f_2$ in Eqs. (18,19) by the ones of the $f'_2$, 
the corresponding decay widths for the $f'_2$ are determined. 
\begin{equation}
\frac{\Gamma(f'_2\rightarrow K\bar{K})}{\Gamma(f_2\rightarrow K\bar{K})} = 8.59
\end{equation}
is obtained.
The data[7] is $7.63(1 \pm 0.23)$. Theory agrees with data within the experimental errors. Now the same $|T|^2$ is used to calculate
the decay widths of $\Gamma(G_2\rightarrow\pi\pi,\;K\bar{K},\;\eta\eta)$. The formulas of the decay widths are obtained  
\begin{eqnarray}
\Gamma(G_2\rightarrow\pi\pi) = |T|^2 m_{G_2} (a_3 + b_3)^2(1-{4m^2_\pi\over m^2_{G_2}})^{{5\over2}},\\
\Gamma(G_2\rightarrow K\bar{K}) = |T|^2 m_{G_2} {1\over3}(2 b_3 - a_3))^2(1-{4m^2_K\over m^2_{G_2}})^{{5\over2}},\\
\Gamma(G_2\rightarrow\eta\eta) = |T|^2 m_{G_2} {1\over3}(b_3 - a_3))^2(1-{4m^2_{\eta}\over m^2_{G_2}})^{{5\over2}}.
\end{eqnarray}
The $|T|^2$ is determined by $\Gamma(f_2\rightarrow\pi\pi) = 158.1 (1 \pm 0.04) \textrm{MeV}$ [7]. The numerical results are
\begin{eqnarray}
\Gamma(G_2\rightarrow\pi\pi) = 189 (1 \pm 0.04) \textrm{MeV},\;\Gamma(G_2\rightarrow K\bar{K}) = 0.23 (1 \pm 0.04) \textrm{MeV},
\nonumber \\
\Gamma(G_2\rightarrow\eta\eta) = 1.82 (1 \pm 0.04)\textrm{MeV}.
\end{eqnarray}
The total decay width of the $G_2(1429)$ is 191 $\textrm{MeV}$. Because the coefficients, $2 b_3 - a_3$, $b_3 - a_3$,
and the phase space of the $K\bar{K}$ and $\eta\eta$ channels are smaller the decay modes of $K\bar{K}$ and $\eta\eta$ are strongly suppressed. 
The $\pi\pi$ decay mode
dominates the hadronic decays of the $G_2$ state. In Ref. [11] a $f_2(1430)$ state with the width of $150 \pm 50\textrm{MeV}$
has been reported. 

\section{The decay $J/\psi\rightarrow\gamma G_2$}
In QCD the radiative decay of the $J/\psi$ is described as $J/\psi\rightarrow\gamma gg$. Therefore, a state with larger glueball
component should have larger production rate in $J/\psi$ radiative decay.
Using the mixing (15-17), we can study the branching ratios of $J/\psi\rightarrow\gamma f_2,\;
\gamma f'_2,\;\gamma G_2$. In Ref. [4] the expression of the decay width of $J/\psi\rightarrow \gamma g_2$  is presented
\begin{equation}
\Gamma(J/\psi\rightarrow\gamma g_2)=\frac{128\pi\alpha}{81}\alpha_s^2(m_c)G^2(0)\psi^2_{J}(0) c^2
{1\over m^4_c}
(1-{m^2\over m^2_{J}})\{T^2_0+T^2_1+T^2_2\},
\end{equation}
where $\psi_{J}(0)$ is the wave functions of $J/\psi$ at origin, G(0) is a parameter related to the $2^{++}$ glueball $g_2$ [4],
m is the mass of the physical state whose branching ratio is going to calculate, and c is the coefficient of the glueball 
component of the states of $f_2,\;f'_2$ and $g_2$ respectively.  
\begin{eqnarray}
T_0=-{2\over\sqrt{6}}(A_2+p^2 A_1),\;\;
T_1=-{\sqrt{2}\over m_{J}}(E A_2+m p^2 A_3),\;\;
T_2=-2A_2,\nonumber\\
E={1\over2m}(m^2_{J}+m^2),\;\;p={1\over2m}(m^2_{J}-m^2),\nonumber\\
A_1=-a\frac{2m^2-m_{J}(m_{J}-2m_c)}{m_c m_{J}[m^2_c+{1\over4}(m^2_{J}-2m^2)]},\;\;
A_2=-a{1\over m_c}\{{m^2\over m_{J}}-m_{J}+2m_c\},\nonumber \\
A_3=-a\frac{m^2-{1\over2}(m_{J}-2m_c)^2}{m_c m_{J}[m^2_c+{1\over4}(m^2_{J}-2m^2)]},\;\;
a={16\pi\over 3\sqrt{3}}{\sqrt{m_{J}}\over m^2_c}.
\end{eqnarray}
Replacing $m_c$, $m_J$, and $Q_c$ by $m_b$, $m_{\Upsilon}$, and $Q_b$ respectively in Eqs. (29-30),
the decay $\Upsilon(1S)\rightarrow\gamma f_2$ is studied[12]. Theory agrees with data very well. 
 
The ratios of the helicity amplitudes of the $J/\psi\rightarrow\gamma f_2$
\[x = {T_1\over T_0},\;\;y = {T_2\over T_0}, \]
have been measured. The early measurements show
\[x = 0.88 \pm 0.11,\;y=0.04 \pm 0.14\;\; Crystal Ball [13],\]
\[x = 0.81 \pm 0.16,\;y=0.02 \pm 0.15\;\; MarK II [13],\]
\[x = 0.6 \pm 0.3,\;y=0.3 \pm 0.6\;\; Pluto [13].\]
The BES Collaboration has reported following results [14]
\[x = 0.89 \pm 0.02 \pm 0.10,\;y = 0.46 \pm 0.02 \pm 0.19.\]
The values of x obtained by BES [14] is consistent with other measurements [13]. However, the value of y 
obtained by BES Collaboration is much larger. As pointed in Ref. [4], the value of y and the decay rate (29,30) 
are very sensitive to the value of $m_c$. In Ref. [4] small $y = 0.04$ and $x = 0.66$ are obtained by taking 
$m_c = 1.3 \textrm{GeV}$. 

The branching ratios of $J/\psi\rightarrow\gamma f_2,\;\gamma f'_2$ are measured [7]
\[B(J/\psi\rightarrow\gamma f_2) = (1.43 \pm 0.11)\times 10^{-3},\;
B(J/\psi\rightarrow\gamma f'_2) = (4.5 ^{+0.7}_{-0.4})\times 10^{-4}.\]
Using the Eqs. (29,30) and choosing the value of the $m_c$ we obtain
\begin{enumerate}
\item Taking $m_c = 1.3 \textrm{GeV}$ and inputting $B(J/\psi\rightarrow\gamma f_2)$, we obtain\\
for the decay $J/\psi\rightarrow\gamma f_2$ 
\[x = 0.66,\;\;y = 0.04;\]
for the decay $J/\psi\rightarrow\gamma f'_2$
\[x = 0.79,\;y=0.28;\]
for the decay $J/\psi\rightarrow\gamma G_2$
\[x=0.74,\;y=0.20;\]

\begin{eqnarray}
B(J/\psi\rightarrow\gamma f'_2) = 1.04 (1 \pm 0.08)\times 10^{-3},\nonumber \\
B(J/\psi\rightarrow\gamma G_2) = 2.45 (1 \pm 0.08)\times 10^{-3}.
\end{eqnarray}
The branching ratio of $J/\psi\rightarrow\gamma f'_2$ is greater than the data $(4.5^{+0.7}_{-0.4})\times 10^{-4}$ [6].
\item Taking $m_c = 1.5\textrm{GeV}$ and inputting $B(J/\psi\rightarrow\gamma f_2)$, we obtain\\
for the decay $J/\psi\rightarrow\gamma f_2$
\[x = 0.7,\;y=0.37, \]
they are consistent with the data of Ref. [14];

for the decay $J/\psi\rightarrow\gamma f'_2$
\[x = 0.84,\;y=0.55;\]
for the decay $J/\psi\rightarrow\gamma G_2$
\[x=0.79,\;y=0.47;\]

\begin{eqnarray}
B(J/\psi\rightarrow\gamma f'_2) = 0.67 (1 \pm 0.08)\times 10^{-3},\nonumber \\
B(J/\psi\rightarrow\gamma G_2) = 1.77 (1 \pm 0.08)\times 10^{-3}.
\end{eqnarray}
The branching ratio of $J/\psi\rightarrow\gamma f'_2$ is closer to the data [6]. There are other two measurements
\[B(J/\psi\rightarrow\gamma f'_2) = (5.6 \pm 1.4 \pm 0.9)\times 10^{-3} [15],\;\; (6.8 \pm 1.6 \pm 1.4)\times 10^{-3} [16].\]
\end{enumerate}
The values of  x and y of $J/\psi\rightarrow\gamma f_2$ and $B(J/\psi\rightarrow\gamma f'_2)$ obtained favors $m_c = 1.5\textrm{GeV}$. 

In both cases the $B(J/\psi\rightarrow\gamma G_2)$ predicted is greater than $B(J/\psi\rightarrow\gamma f_2)$.
\section{Summary}
Using the argument of the $N_C$ and chiral expansions, the mixing between $f_8,\;f_0$, and a $2^{++}$
glueball state is determined. The mass of a new $2^{++}$ state $G_2$ is predicted. The predicted branching ratios of $f_2\rightarrow
K\bar{K},\;\eta\eta$ and $f'_2\rightarrow\eta\eta$ agree with data reasonably well.
The hadronic decay widths of the new state $G_2$ are predicted. The $\pi\pi$ decay mode is dominant and the $K\bar{K}$ and 
the $\eta\eta$ modes are strongly suppressed. The predicted branching ratio of $J/\psi\rightarrow\gamma G_2$ is larger
than the branching ratio of $J/\psi\rightarrow\gamma f_2$.

\end{document}